\documentclass[12pt]{article}

\usepackage{graphicx}

\def\rf#1{(\ref{eq:#1})}
\def\lab#1{\label{eq:#1}}

\def\br{\begin{eqnarray}}
\def\er{\end{eqnarray}}
\def\be{\begin{equation}}
\def\ee{\end{equation}}
\def\({\left(}
\def\){\right)}

\def\u2{\mid u\mid^2}
\def\rlx{\relax\leavevmode}
\def\IR{\rlx\hbox{\rm I\kern-.18em R}}

\begin{document}

\begin{titlepage}
\vspace*{-1cm}

\vskip 3cm

\vspace{.2in}
\begin{center}
{\large\bf Spinning Hopf solitons on $S^3\times \IR$}
\end{center}

\vspace{.5cm}

\begin{center}
A.C. Ris\'erio do Bonfim~$^{2}$ and L. A. Ferreira~$^{1,2}$

\vspace{.5 in}
\small

\par \vskip .2in \noindent
$^{(1)}$Instituto de F\'\i sica de S\~ao Carlos; IFSC/USP;\\
Universidade de S\~ao Paulo  \\ 
Caixa Postal 369, CEP 13560-970, S\~ao Carlos-SP, Brazil\\

\par \vskip .2in \noindent
$^{(2)}$~Instituto de F\'\i sica Te\'orica - IFT/UNESP\\
Universidade Estadual Paulista\\
Rua Pamplona 145\\
01405-900,  S\~ao Paulo-SP, Brazil\\

\normalsize
\end{center}

\vspace{.5in}

\begin{abstract}

We consider a field theory with target space being the two dimensional
sphere $S^2$ and defined on the space-time $S^3\times \IR$. The
Lagrangean is the square of the pull-back of the area form on
$S^2$. It is invariant under the conformal group $SO(4,2)$ and the
infinite dimensional group of area preserving diffeomorphisms of
$S^2$. We construct an infinite number of exact soliton solutions with
non-trivial Hopf topological charges. The solutions spin with a
frequency which is bounded above by a quantity proportional to the
inverse of the radius of $S^3$. The construction of the
solutions is made possible by an ansatz which explores the conformal
symmetry and a $U(1)$ subgroup of the area preserving diffeomorphism
group.

\end{abstract} 
\end{titlepage}

\section{Introduction}

The development of exact methods is very important for the
understanding of several 
non-perturbative aspects of field theories of physical
interest. Soliton solutions are in one way or the other connected to
the few advances in this area. Since the appearance of such special
type of solutions require a large degree of symmetries, the
integrability properties play a fundamental role. In low dimensions
the structures responsible for the integrability and soliton solutions
is reasonably well understood. In higher dimensions, although some
exact results have been obtained,  the
essence of the structures involving solitons in such field theories  is 
far from clear. For gauge theories the most important and relevant
results were obtained using supersymmetry and duality concepts
\cite{duality}. However, for general Lorentz invariant field theories
in dimensions higher than two, mathematical concepts like loop spaces
may play an important role as well, specially in the classical
integrability structures \cite{afsg}. The study of models that present
exact soliton solutions is very important as a laboratory to test
ideas and methods.  

In this paper we consider a non-linear field theory on the space-time
$S^3\times \IR$,  with the fields taking values on the sphere
$S^2$, and the Lagrangean being the pull-back of the area form on the target
space. The theory is integrable in the sense of possessing a zero
curvature representation within the approach of \cite{afsg}, an
infinite number of conserved currents, and exact soliton
solutions. From the topological point of view it belongs to a class of
models, with target space $S^2$, presenting  solitons
with non-trivial Hopf topological charges, and where the best known example is
the Skyrme-Faddeev model \cite{fn,ward}.  

We construct an infinite number of exact soliton solutions with finite
energy, non-trivial Hopf topological charges, and that spin with a
frequency $\omega$ that is bounded above by a quantity inversely
proportional to  the radius $r_0$ of $S^3$. Our work
complements that of ref. \cite{carli}  where time dependent solitons
for the same model were also constructed. In addition,  our
solutions present properties similar to those of the solitons 
constructed in \cite{timehopf} for a theory with the same Lagrangean but
on Minkowski space-time. 

The action of the model we consider is given by 
\be
S= -\frac{1}{e^2}\, \int dt\; \int_{S^3} d\Sigma  \; H^2_{\mu\nu}
\lab{action}
\ee
where $H_{\mu\nu}$ is the pull-back of the area form on $S^2$ given by
\be
H_{\mu\nu}\equiv 
-2i\frac{\(\partial_{\mu} u\partial_{\nu} u^* - 
 \partial_{\nu} u \partial_{\mu} u^*\)}{\(1+\u2\)^2}  
=  {\vec n}\cdot\(\partial_{\mu}{\vec n} \wedge 
\partial_{\nu}{\vec n}\)
\lab{hdef} 
\ee
where $u$ is a complex scalar field, and ${\vec n}$ (${\vec n}^2=1$)
is a triplet of 
real scalar fields living on $S^2$. They are related by the
stereographic projection
\be
{\vec n} = \frac{1}{1+\u2}\, \(u+u^*,-i\(u-u^*\),\u2 -1\)
\lab{ndef}
\ee
The Euler-Lagrange equations associated to \rf{action} are given by 
\be
\partial_{\mu} {\cal K}^{\mu}=0 
\lab{eqmot}
\ee
and its complex conjugate, and where 
\be
{\cal K}_{\mu} = H_{\mu\nu}\partial^{\nu}u
\lab{kdef}
\ee

The action \rf{action} and equations of motion \rf{eqmot} are invariant under
the group $SO(4,2)$ of 
conformal transformations on $S^3\times \IR$. They are also invariant
under the infinite group of area preserving diffeomorphisms of the
target space $S^2$. The associated Noether currents are given by
\cite{fr} 
\be
J_{\mu}^G = \frac{\delta G}{\delta u} {\cal K}_{\mu} +   
\frac{\delta G}{\delta u^*} {\cal K}_{\mu}^*
\lab{noethercurr}
\ee
with $G$ being any functional of $u$ and $u^*$, but not of their
derivatives. The conservation of the currents \rf{noethercurr} is a
consequence of the equations of motion \rf{eqmot}, and the fact that
\be
{\cal K}_{\mu}\partial^{\mu}u=0\qquad\qquad 
 {\cal K}_{\mu}\partial^{\mu}u^* +{\cal K}_{\mu}^*\partial^{\mu}u=0
\ee

The model \rf{action} has been considered in ref. \cite{carli} where
an infinite set of soliton solutions, static and time dependent, with
non-trivial Hopf charges have been constructed. The basic ingredient
of the construction in \cite{carli} was the use of special coordinates
on $S^3\times \IR$ which in its turn leads to a powerful ansatz. The
group of area preserving diffeomorphisms has a $U(1)_{\alpha}$ subgroup
generated by the phase transformations $u \rightarrow e^{i\alpha}\,
u$, with $\alpha$ constant. Considering commuting $U(1)_{\varphi_i}$'s
subgroups 
in the conformal group one build an ansatz where the  field
configurations are 
invariant under each of the diagonal subgroups of
$U(1)_{\alpha}\otimes U(1)_{\varphi_i}$ \cite{bf}. The special
coordinates are then chosen in such a way that the  transformations
associated to the $U(1)_{\varphi_i}$'s correspond to translation along
them. However, for the method to work, the subgroups
$U(1)_{\varphi_i}$'s do not have to be compact as $U(1)_{\alpha}$ is,  and
so translations along non-compact directions, as time for instance, can
be included into the ansatz. A more detailed account of these methods,
valid on any space-time, will be given in ref. \cite{bonfim}. 

The paper is organized as follow: in sec. \ref{sec:solutions} we
present the special coordinates and the corresponding ansatz, which
reduces the four dimensional non-linear  partial differential
equations \rf{eqmot} into a single linear ordinary differential equation for a
real profile function. The solutions with appropriated boundary
conditions is then constructed. In sec. \ref{sec:hopf} we calculate
the Hopf topological charges for those solutions. The Noether charges
associated to the area preserving diffeomorphisms of $S^2$ are
calculated in sec \ref{sec:noether}. The angular momenta associated to
the $SO(4)$ rotations on $S^3$ are evaluated in
sec. \ref{sec:rotations}, and the energies of the solutions are given in
sec. \ref{sec:energy}. We give an appendix with the details of the
calculations involved in the evaluation of the Noether charges. 

\section{Soliton solutions}
\label{sec:solutions}

Following \cite{carli} we then introduce a set of coordinates in the
space-time $S^3\times \IR$ such that the metric is given by 
\be
ds^2=  c^2\, dt^2 - r_0^2\, \( \frac{dz^2}{4\, z\(1-z\)}+\(1-z\)
d{\varphi_1}^2+  z\, d{\varphi_2}^2 \)
\lab{metric}
\ee
where $t$ is the time, $c$ is the speed of light, $z$ and $\varphi_i$,
$i=1,2$, are coordinates on the sphere $S^3$, and $0\leq z \leq 1$,
$0\leq \varphi_i \leq 2 \pi$, and $r_0$ is the radius of the sphere
$S^3$. Embedding $S^3$ on $\IR^4$ we get that
the Cartesian coordinates of the points of $S^3$ are \br
x_1&=& r_0\, \sqrt{z}\, \cos \varphi_2 \qquad \qquad 
x_3= r_0\,\sqrt{1-z}\, \cos \varphi_1 \nonumber\\  
x_2&=& r_0\,  \sqrt{z}\, \sin \varphi_2 \qquad \qquad \,
x_4=  r_0\,\sqrt{1-z}\, \sin \varphi_1
\lab{cartcoords3}
\er 
The ansatz is given by \cite{bf,carli,afz99,bonfim}\footnote{On the
  completion of this paper we became aware of ref. \cite{asgw} where
  that ansatz, with a different profile function, was used on some
  other models on $S^3\times \IR$.}
\be
u = \sqrt{\frac{1-g}{g}}\,  e^{i\( m_1 \varphi_1 + m_2
  \varphi_2+\omega t\)}
\lab{ansatz}
\ee
where $m_i$, $i=1,2$, are arbitrary integers, $\omega$ is a real
frequency, and $g$ is a real profile function. In addition, we require
that $0\leq g \leq 1$ for the factor $\sqrt{\(1-g\)/g}$ to be
real. The form of that factor is important because it implies $g$
satisfies a linear ordinary differential equation. Indeed, replacing
\rf{ansatz} into \rf{eqmot} one gets 
\be
\partial_z\(\Omega \,\partial_z \, g\)=0
\lab{eqg}
\ee
with
\be
\Omega = m_1^2 \, z+m_2^2 \, \(1-z\) - \frac{r_0^2\omega^2}{c^2} \,
z\(1-z\) 
\lab{omegadef}
\ee
Since we have to satisfy the condition $0\leq g \leq 1$, we can not
have zeroes of $\Omega$ lying in the
allowed range for $z$, i.e., $0\leq z \leq 1$. 

For $\omega=0$ one has
that the zeroes of $\Omega$ are $z=0$ for $m_2=0$, $z=1$ for
$m_1=0$, and they lie in the intervals $z>1$ or $z<0$, for non-vanishing
$m_1$ and $m_2$. Therefore, we have to discard the cases $m_1=0$ or
$m_2=0$. The solutions
satisfying the boundary conditions $g\(0\)=0$ and $g\(1\)=1$, for
$\omega=0$   are then   
\be
g=z\; ; \qquad\qquad\qquad \qquad {\rm for} \qquad \omega=0 \quad
m_1^2 = m_2^2
\lab{static1}
\ee
and
\be
g=\frac{\ln\(\(q^2-1\)\, z+1\)}{\ln q^2}\; ;\qquad \qquad {\rm for}
\qquad\omega=0 \quad  m_1^2\neq m_2^2
\lab{static2}
\ee
where 
\be
q\equiv \frac{\mid m_1\mid}{\mid m_2\mid}
\lab{qdef}
\ee
These are the static solutions of \cite{carli}.  

For $\omega\neq 0$ we  write $\Omega
=\frac{r_0^2\omega^2}{c^2} \,\(z-z_{+}\)\(z-z_{-}\)$, with
$z_{\pm}=-b\pm\sqrt{\Delta}$, and where
\be
b=\frac{1}{2}\, \(p_{+} \, p_{-} - 1\)\; ; \qquad  
\Delta=\frac{1}{4}\, \(p_{+}^2-1\)\(p_{-}^2-1\)\; ;\qquad  
p_{\pm}\equiv \frac{c}{r_0\,\omega}\(m_1\pm m_2\)
\ee
Notice that $b$ and $\Delta$ are invariant under the independent change
of signs of $m_1$, $m_2$ and $\omega$. 
The solutions for \rf{eqg} are $g\sim \frac{1}{z-z_{+}} +{\rm const.}$, for
$z_{+}=z_{-}$, and $g\sim \ln \frac{z-z_{+}}{z-z_{-}} +{\rm const.}$
for $z_{+}\neq z_{-}$. Therefore, we can not have one or both of the
zeroes $z_{\pm}$ of $\Omega$, lying in the interval $\left[0,1\right]$,
since $g$ will have divergencies and so the condition $0\leq g\leq 1$
will not be satisfied. A careful analysis shows that $z_{\pm} \in
\left[0,1\right]$, when both $p_{\pm}$ lie in the interval
$\left[-1,1\right]$, and so when $\frac{r_0^2\,\omega^2}{c^2}\geq
\(\mid m_1\mid +\mid m_2\mid\)^2$. In
addition, if $p_{\pm}^2>1$, then $z_{+}=0$ for $p_{+}=p_{-}$, and
$z_{-}=1$ for $p_{+}=-p_{-}$. Therefore, we can not have $m_1=0$ or
$m_2=0$. That is also true in the case $\omega=0$, and consequently,
as we show below, it implies that all our solutions will have
non-trivial Hopf topological charges. 

Therefore the solutions for $\omega \neq 0$, satisfying the boundary
conditions $g\(0\)=0$ and $g\(1\)=1$, are
\begin{enumerate}
\item In the cases where $p_{\pm}^2>1$ ($\Delta >0$) one has 
\be
g=\left[\ln\frac{z+b-\sqrt{\Delta}}{z+b+\sqrt{\Delta}}-
\ln\frac{b-\sqrt{\Delta}}{b+\sqrt{\Delta}}\right]/
\ln\frac{a+\sqrt{\Delta}}{a-\sqrt{\Delta}}\; ; 
\qquad 
\frac{r_0^2\,\omega^2}{c^2}<\(\mid m_1\mid-\mid m_2\mid\)^2
\lab{solg1}
\ee
with $a=\frac{1}{2}\(\frac{p_{+}^2+p_{-}^2}{2}-1\)$. 
\item In the cases where $p_{+}^2=1$ and $p_{-}^2>1$, or $p_{-}^2=1$ and
  $p_{+}^2>1$, ($\Delta =0$) one has 
\be
g=\frac{q}{q-1}\( 1 - 
\frac{1}{\(q -1\)\, z + 1}\)\; ; \qquad \qquad 
\frac{r_0^2\,\omega^2}{c^2}=\(\mid m_1\mid-\mid m_2\mid\)^2
\lab{solg2}
\ee
with $q$ given in \rf{qdef}. 
\item In the cases where $p_{+}^2<1$ and $p_{-}^2>1$, or $p_{+}^2>1$
  and $p_{-}^2<1$, ($\Delta <0$) one has 
\br
g=\left[ {\rm ArcTan}\frac{\sqrt{-\Delta}}{b}\right. &-& \left. 
{\rm ArcTan}\frac{\sqrt{-\Delta}}{b+z}\right]/
{\rm ArcTan}\frac{\sqrt{-\Delta}}{a}
\lab{solg3}\\
&&  \qquad 
\(\mid m_1\mid-\mid m_2\mid\)^2<\frac{r_0^2\,\omega^2}{c^2}<
\(\mid m_1\mid+\mid m_2\mid\)^2 \nonumber 
\er
and where the ${\rm ArcTan}$ takes values between $0$ and $\pi$. 
\end{enumerate}
In all those cases $g$ is a monotonically increasing function of $z$
starting with $g=0$ at $z=0$ and finishing with $g=1$ at $z=1$. Notice
that given the values of $m_1$ and $m_2$, the type of profile function
$g$  is determined by the dimensionless quantity
$\frac{r_0\,\omega}{c}$. In addition, the range in which such 
dimensionless quantity  can vary is restricted by those same
integers. Therefore, the frequency of rotation of the solution is
inversely proportional to the radius $r_0$ of the sphere $S^3$. We
point out that the period, $\frac{2\pi}{\omega}$, of such rotations
is never shorter than the time that a light ray takes to travel along a
maximum circle on $S^3$. 

In all the cases in \rf{solg1}-\rf{solg3}, we have that
\be
\partial_z\, g = \frac{\beta}{\Omega}  
\lab{derg}
\ee
with
\be
\beta\equiv \(
m_1^2+m_2^2-\frac{r_0^2\,\omega^2}{c^2}\)\frac{w}{\ln\frac{1+w}{1-w}}
\; ; \qquad\qquad \qquad \qquad 
w\equiv \frac{\sqrt{\Delta}}{a}
\lab{betawdef}
\ee
In the case where $\Delta =0$, one has that $\beta$ simplifies to
$\beta=\mid m_1\,m_2\mid$.

The form of the solution can be visualized by the surfaces in $S^3$ of
constant $n_3$, the third component of ${\vec n}\in S^2$, given in \rf{ndef}. 
Notice that $n_3$ depends on $\mid u\mid^2$ which in its turn depends
on $g$ (see \rf{ansatz}). But since $g$ is a monotonic function of
$z$, it implies that constant $n_3$ means constant $z$. Therefore, the
surfaces of constant $n_3$ can be obtained from \rf{cartcoords3} by
fixing $z$ and varying $\varphi_1$ and $\varphi_2$. Such surfaces are
the same for all solutions. What changes from one solution to the
other is the correspondence between $z$ and $n_3$, determined through
\rf{solg1}-\rf{solg3} by the triple $\(m_1,m_2,\omega\)$. Notice that such
surfaces do not evolve in time, since $\mid u\mid^2$ is time
independent. For all solutions,
the pre-image of the north pole of $S^2$, ${\vec n}=\(0,0,1\)$
(and so $z=0$), corresponds to a circle of radius $r_0$ on the plane
$x^3x^4$, and the pre-image of the south pole of $S^2$, ${\vec n}=\(0,0,-1\)$
(and so $z=1$), corresponds to a circle of radius $r_0$ on the plane
$x^1x^2$. The pre-image of a given ${\vec n}$ with $-1<n_3<1$, is a two
dimensional surface of a torus like shape. 

\section{The Hopf topological charge}
\label{sec:hopf}

At any fixed time $t$ our solution defines a map from the physical
space $S^3$ to the target space $S^2$, and so it is a Hopf map. The
Hopf invariant, or linking number, is calculated as follow. First we
define a map from the physical $S^3$ to another $3$-sphere $S^3_Z$, as
\br
Z=\(
\begin{array}{c}
w_1\\
w_2
\end{array}\) = 
\(
\begin{array}{c}
\sqrt{1-g} \, e^{i\, \(m_1 \varphi_1+\omega \, t\)}\\
\sqrt{g} \, e^{-i \, m_2 \varphi_2}
\end{array}\) 
\er
where $w_1$ and $w_2$ are two complex coordinates parametrizing
$S^3_Z$, such that $Z^{\dagger}\; Z= \mid w_1\mid^2+\mid
w_2\mid^2=1$. Then we map $S^3_Z$ into the target space $S^2$ as
$u=w_1/w_2$. The Hopf invariant is defined through the integral
\be
Q_H= \frac{1}{4\pi^2}\int_{S^3} d \Sigma \, {\vec A}\cdot \({\vec \nabla}
\wedge {\vec A}\) 
\ee
where ${\vec \nabla}$ is gradient  on the physical $S^3$, 
\be
d \Sigma =\frac{r_0^3}{2}\, dz\, d\varphi_1\, d\varphi_2
\lab{volume}
\ee
is the volume element on  $S^3$, and 
\be
{\vec A}= \frac{i}{2}\, \( Z^{\dagger}\;{\vec \nabla} Z - {\vec
  \nabla} Z^{\dagger}\; Z\)
\ee
Evaluating it one gets \cite{carli}
\be
{\vec A}= -\frac{m_1}{r_0}\frac{\(1-g\)}{\sqrt{1-z}}{\hat {\bf
    e}}_{\varphi_1} +  
\frac{m_2}{r_0}\frac{g}{\sqrt{z}}{\hat {\bf e}}_{\varphi_2} \; ;\qquad 
{\vec \nabla} \wedge {\vec A} = 
\frac{2}{r_0^2} \partial_z\,g\(-m_2\sqrt{1-z}{\hat {\bf e}}_{\varphi_1}
+m_1\sqrt{z}{\hat {\bf e}}_{\varphi_2} \)
\ee
and so, ${\vec A}$ is time independent. Therefore
\be
Q_H= m_1\, m_2\(g\(1\)-g\(0\)\)= m_1\, m_2
\lab{hopffinal}
\ee
That is valid for all solutions \rf{ansatz} and \rf{solg1}-\rf{solg3},
since they all satisfy the boundary condition $g\(1\)=1$ and
$g\(0\)=0$. Since, from our considerations in
sec. \ref{sec:solutions}, the integers $m_1$ and $m_2$ can never vanish,
all solutions have non-trivial Hopf topological charges. 

\section{The Noether charges}
\label{sec:noether}

One can check that the time component of the vector ${\cal K}_{\mu}$,
defined in \rf{kdef}, evaluated on the ansatz configurations
\rf{ansatz} is given by
\be
{\cal K}_0=-\frac{4}{r_0^2}\, \frac{\omega}{c} \,
\frac{z\(1-z\)}{g\(1-g\)} \, \(\partial_z\, g\)^2 \; u 
\lab{k0}
\ee
Therefore, the density of the Noether charges associated to
\rf{noethercurr}, with a functional of the form $G = u^m\, {u^*}^n$ is
given by ($m$ and $n$ are integers for $G$ to be single valued for any
configuration \rf{ansatz})
\be
J_0^{(m,n)}= -\(m+n\)\frac{4}{r_0^2}\, \frac{\omega}{c} \,
\frac{z\(1-z\)}{g\(1-g\)} \, \(\partial_z\, g\)^2 \; 
\(\frac{1-g}{g}\)^{\frac{(m+n)}{2}}\,
e^{i\(m-n\)\(m_1\varphi_1+m_2\varphi_2+\omega t\)}
\ee
So, if $m\neq n$, the corresponding charge will vanish 
since the integral of the phase factor with the volume element
\rf{volume} is zero. Consequently, the relevant Noether charges are
those where $G$ is a functional of the norm of $u$, or equivalently a
functional of $g$. Such infinite set of charges have vanishing Poisson
brackets \cite{fr}, and so should be important for the integrability
of the model. In fact, such abelian subalgebra was shown to be
connected to constraints leading to integrable submodels of some theories
with target space $S^2$ \cite{joaqconst}. 
Using \rf{noethercurr} and \rf{k0} one then gets that, for
$G\equiv G\(g\)$, 
\be
Q^G = \int_{S^3} d\Sigma \, J_0^G= 16\,\pi^2\,\frac{r_0\,\omega}{c}\,
\int_0^1 dz\, z\(1-z\)\,\(\partial_z\, g\)^2\, \frac{\delta G}{\delta
  g}
\lab{noetherg}
\ee
Choosing $G=g^{n}/16\pi^2 n!$, with $n$ a positive integer, we
get that
\be
Q^{(n)}= \frac{r_0\,\omega}{c}\, F^{(n)}\(w\) \qquad \qquad \qquad n=1,2,3
\ldots
\lab{noethern}
\ee
where
\br
 F^{(n)}\(w\)\equiv \frac{1}{\(\ln\frac{1+w}{1-w}\)^{n+1}}\left[ 
-2\, \epsilon_{-}\(n\) + 
\sum_{l=1}^n\(\frac{\epsilon_{+}\(n-l\)}{w}-\epsilon_{-}\(n-l\)\)\frac{1}{l!}\,
\(\ln\frac{1+w}{1-w}\)^l\right] 
\lab{fndef}
\er
with $\epsilon_{\pm}\(n\)\equiv \(1\pm\(-1\)^n\)/2$, and $w$ defined
in \rf{betawdef}. In appendix \ref{app:a} 
 we give the details of the calculations leading to
\rf{noethern}. In the case where $\Delta =0$, we have that the Noether
charges simplify to $Q^{(n)}=
\frac{r_0\,\omega}{c}\, \frac{n}{\(n+2\)!}$.

\section{The angular momentum}
\label{sec:rotations}

The action \rf{action} is invariant under the conformal group
$SO(4,2)$ of the space-time $S^3\times \IR$. Such group contains the 
subgroup $SO(4)$ of rotations on the sphere $S^3$. In
ref. \cite{bonfim} we shall present a more detailed study of how the
$SO(4,2)$ conformal transformations, and $SO(4)$ rotations, are
realized in terms of the coordinates
\rf{metric}-\rf{cartcoords3}. Here we present the six Noether charges
associated to the $SO(4)$ symmetry. According to \rf{cartcoords3}, the
rotations on the planes 
$x^1x^2$ and $x^3x^4$ correspond to the translations on the angles
$\varphi_2$ and $\varphi_1$ respectively. The  density of the
corresponding Noether charges, for the ansatz configurations
\rf{ansatz}, are
\be
J^0_{\varphi_i} = \frac{64\, \omega}{e^2\,r_0^2\, c} \, m_i \, z\(1-z\)\,
\(\partial_z\, g\)^2 \qquad \qquad\qquad \qquad i=1,2
\ee
Integrating it with the measure \rf{volume}, for the solutions
\rf{solg1}-\rf{solg3},  one gets the charges 
\be
Q_{\varphi_i} = \frac{128\, \pi^2}{e^2} \, \frac{r_0\,\omega}{c}\, m_i
\, F^{(1)}\(w\)\qquad \qquad\qquad \qquad i=1,2
\lab{angmom}
\ee
with $F^{(1)}\(w\)$ as in \rf{fndef}, i.e.
\be
F^{(1)}\(w\) =  
\frac{1}{w\, \ln\frac{1+w}{1-w}} 
-\frac{2}{\(\ln\frac{1+w}{1-w}\)^2} 
\lab{f1def}
\ee
The density of the Noether charges, for the ansatz configurations
\rf{ansatz}, associated to the rotations on the
remaining four planes $x^ix^j$, are given by
\br
J^0_{\delta_1\delta_2}&=& \frac{64\, \omega}{e^2\,r_0^2\, c} \, z\(1-z\)\,
\(\partial_z\, g\)^2\left[ - m_1\, \sqrt{\frac{z}{1-z}}\,
  \sin\(\varphi_1+\delta_1\)\, \cos\(\varphi_2+\delta_2\)\right.\nonumber\\
 &+&\left. m_2\, \sqrt{\frac{1-z}{z}}\,\cos\(\varphi_1+\delta_1\)
  \sin\(\varphi_2+\delta_2\) \right]
\er
with $\delta_i=0,\frac{\pi}{2}$, $i=1,2$. Therefore, the corresponding Noether
charges vanish since the densities are periodic in the angles
$\varphi_i$.

\section{The Energy}
\label{sec:energy}

The Hamiltonian density associated to \rf{action}, evaluated on the
ansatz configurations \rf{ansatz}, is given by
\be
{\cal H}= \frac{32}{e^2\,r_0^4}\, \(\partial_z\, g\)^2\left[ m_1^2 \,
  z +m_2^2\, \(1-z\)+\frac{r_0^2\,\omega^2}{c^2}\, z\(1-z\)\right]
\ee
For the static solutions \rf{static1}-\rf{static2} the energy is given
by
\be
E = \int_{S^3} d\Sigma \, {\cal H}= \frac{64\,\pi^2}{e^2\,r_0}\, \mid
m_1 m_2\mid \, \frac{\(q-1/q\)}{\ln q^2}
\lab{staticenergy}
\ee
which is the result of \cite{carli}. In the limit $q\rightarrow 1$
($m_1\rightarrow \pm m_2$), one has 
$E\rightarrow \frac{64\,\pi^2}{e^2\,r_0}\,  m_1^2 $, and so the energy
becomes proportional to the modulus of Hopf charge \rf{hopffinal}.

Using \rf{omegadef}, \rf{derg} and \rf{volume} one gets that the
energy, for the ansatz configurations \rf{ansatz}, 
can be written as
\be
E= \frac{64\,\pi^2}{e^2\,r_0}\,\int_0^1dz  \left[\beta\, \partial_z\, g + 
2 \,\frac{r_0^2\,\omega^2}{c^2}\, z\(1-z\)\(\partial_z\, g\)^2\right]
\ee
Therefore, for the solutions \rf{solg1}-\rf{solg3}, it becomes
\be
E= \frac{64\,\pi^2}{e^2\,r_0}\,  \left[\beta + 
2 \,\frac{r_0^2\,\omega^2}{c^2}\, F^{(1)}\(w\)\right]
\lab{energyfinal}
\ee
with $F^{(1)}\(w\)$ given in \rf{f1def}, and $\beta$ in
\rf{betawdef}. 

In the case where $\Delta =0$, or $\frac{r_0^2\,\omega^2}{c^2}=\(\mid
m_1\mid-\mid m_2\mid\)^2$, one has $\beta = \mid m_1\, m_2\mid$, and
$F^{(1)}\(0\)=1/6$. Then the energy simplifies to 
\be
E= \frac{64\,\pi^2}{3\,e^2\,r_0}\,  
\left(m_1^2+m_2^2+\mid m_1\, m_2\mid \right)
\ee

\begin{figure}
\scalebox{1.45}{
\includegraphics{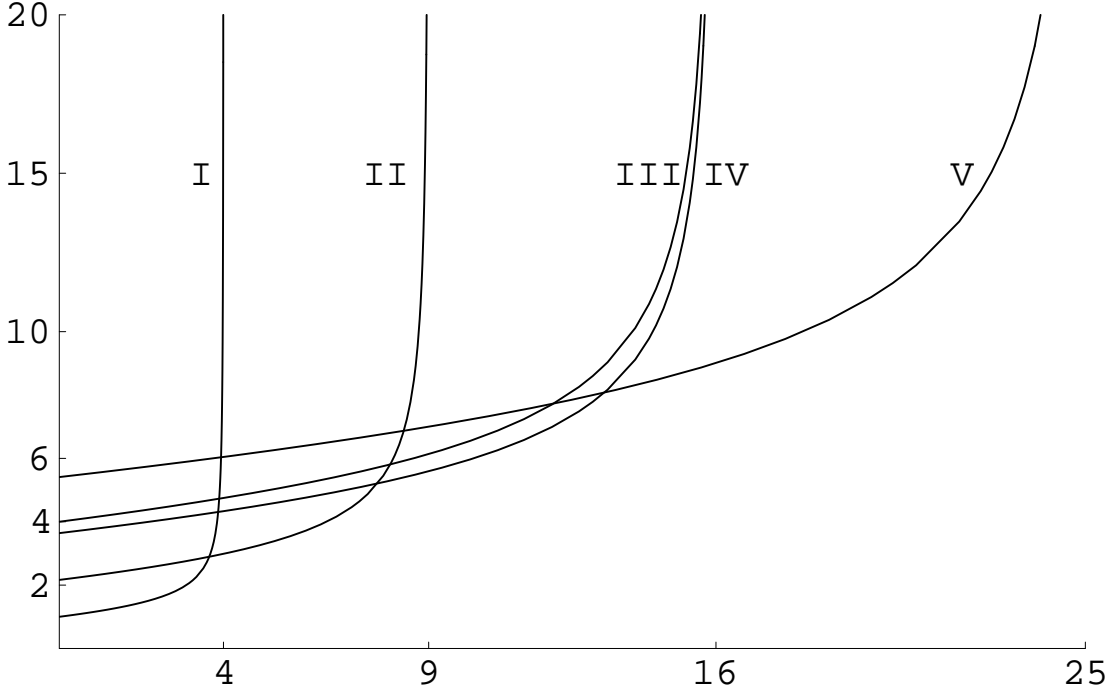}}
\caption{\label{fig:energy} Plot of the energy \rf{energyfinal} (in
  units of $64\,\pi^2/e^2\,r_0$) as a function of
  $r_0^2\,\omega^2/c^2$ for some values of $\(m_1,m_2\)$. The curve I
  corresponds to $\(m_1,m_2\)=\(1,1\)$, curve II to
  $\(m_1,m_2\)=\(1,2\)$, curve  III to $\(m_1,m_2\)=\(1,3\)$, curve IV
  to $\(m_1,m_2\)=\(2,2\)$, and curve V to
  $\(m_1,m_2\)=\(1,4\)$.}   
\end{figure}

The energy \rf{energyfinal} is invariant under the independent change
of the signs of $m_1$, $m_2$ and $\omega$. In addition, it is invariant
under the interchange $m_1\leftrightarrow m_2$. Consequently, the
energy is $16$-fold degenerate for $m_1\neq m_2$ and $\omega\neq 0$,
$8$-fold  degenerate for  $m_1\neq m_2$ and $\omega= 0$, $4$-fold
degenerate for $m_1= m_2$ and $\omega\neq 0$, and $2$-fold
degenerate for $m_1= m_2$ and $\omega= 0$. As we noticed above, there
are no physically acceptable solutions for $m_1=0$ or $m_1=0$. All the
degenerate solutions can be completely distinguished from one another
by the values of the topological Hopf charge \rf{hopffinal}, Noether
charges \rf{noethern}, and angular momenta \rf{angmom}. 

In Figure
\ref{fig:energy} we give the plot of the energy as a function of the
dimensionless quantity $r_0^2\,\omega^2/c^2$ for some values of
$\(m_1,m_2\)$.  Notice that for every solution labelled by the triple
$\(m_1,m_2,\omega\)$, the frequency $\omega$, as discussed in
\rf{solg1}-\rf{solg3}, can vary as $-\(\mid m_1\mid +\mid m_2\mid\)<
\frac{r_0\,\omega}{c}< \mid m_1\mid +\mid m_2\mid$.  The spectrum of
energy of our solutions have three main features: {\em i)} For
$\omega \rightarrow 0$ the energy 
\rf{energyfinal} approaches the static energy \rf{staticenergy}, {\em
  ii)} for fixed $m_1$ and $m_2$ the energy grows monotonically with
$\omega^2$, and {\em iii)} the energy diverges faster, as a function
of $\omega$, for solutions with lower $\mid m_1\mid +\mid m_2\mid$.

\appendix

\section{Appendix: The Noether charge calculations}
\label{app:a}

In order to obtain the Noether charges \rf{noetherg} we have to
evaluate the integral
\be
I^G =\int_0^1 dz\, z\(1-z\)\,\(\partial_z\, g\)^2\, \frac{\delta
  G}{\delta g}
\lab{intgz}
\ee
It is convenient to change the integration variable as
\be
y=\frac{1-z}{z}
\ee
and so
\be
I^G =\int_0^{\infty}dy \, y \,\(\partial_y\, g\)^2\, \frac{\delta
  G}{\delta g} 
\lab{intgy}
\ee
Notice that the integrand of $I^G$ has its form unchanged under the
transformation $y\rightarrow 1/y$, or equivalently $z \rightarrow
1-z$. That is important in relating the Noether charges for solutions
with interchanged boundary conditions at $z=0$ and $z=1$. In
particular, the integral \rf{intgy} is the one appearing in the
expression for the Noether charges of the model considered in
\cite{timehopf}. The
solutions \rf{solg1}-\rf{solg3}, in terms of $y$, are written as 
\be
g=\frac{1}{\ln\frac{y_{-}}{y_{+}}}\, \ln\frac{y-y_{-}}{y-y_{+}}\; ;
\qquad \qquad \qquad\qquad y_{\pm}=\frac{-a\pm\sqrt{\Delta}}{a-b}
\ee
and
\be
\partial_y\, g = -\frac{\beta}{\Lambda}
\ee
where $\beta$ is given in \rf{betawdef}, and 
\be
\Lambda\equiv m_1^2\, \(1+y\)+m_2^2\, y
\(1+y\)-\frac{r_0^2\,\omega^2}{c^2}\, y = m_2^2\(y-y_{+}\)\(y-y_{-}\)
\ee
Therefore, for $G=g^n/16\pi^2n!$, one has that \rf{intgy} becomes
\be
I^G
=\frac{\beta^2}{16\pi^2\(n-1\)!m_2^4}
\frac{1}{\(\ln\frac{y_{-}}{y_{+}}\)^{n-1}}\,   
\int_0^{\infty}dy \, \frac{y}{\(y-y_{+}\)^2\(y-y_{-}\)^2}\, 
\(\ln\frac{y-y_{-}}{y-y_{+}}\)^{n-1}
\ee
One can check that
\be
\frac{d\,k\(n\)}{dy}=\frac{y}{\(y-y_{+}\)^2\(y-y_{-}\)^2}\, 
\(\ln\frac{y-y_{-}}{y-y_{+}}\)^{n-1}
\ee
with
\br
k\(n\)\equiv
-\frac{\(n-1\)!}{\(y_{+}-y_{-}\)^2}&&\left[\frac{y_{-}}{y-y_{-}}
-\(-1\)^{n}\frac{y_{+}}{y-y_{+}}-\frac{y_{+}+y_{-}}{y_{+}-y_{-}}\,
\frac{1}{n!}\,\(\ln\frac{y-y_{-}}{y-y_{+}}\)^{n}\right.\nonumber\\
&+& \left. 
\sum_{l=1}^{n-1}\frac{1}{l!}\,\(\ln\frac{y-y_{-}}{y-y_{+}}\)^{l}\,
P\(n-1-l\)\right]
\er
with
\br
P\(n\)&=& \frac{y^2-y_{+}\,y_{-}}{\(y-y_{+}\)\(y-y_{-}\)} \qquad
\qquad \qquad \qquad \qquad  \quad \mbox{\rm for $n$ even} \\
P\(n\)&=&
-\frac{\(y_{+}+y_{-}\)\(y^2+y_{+}\,y_{-}\)-4\,y\,y_{+}\,y_{-}}{
\(y-y_{+}\)\(y-y_{-}\)\(y_{+}-y_{-}\)} \qquad
\quad \mbox{\rm for $n$ odd} \nonumber
\er
Therefore, we have that
\be
I^G
=\frac{\beta^2}{16\pi^2\(n-1\)!m_2^4}
\frac{1}{\(\ln\frac{y_{-}}{y_{+}}\)^{n-1}}\,
\left[k\(n\)\mid_{y=\infty}-k\(n\)\mid_{y=0}\right] = \frac{1}{16\pi^2}\,
F^{(n)}\(w\)
\ee
with $F^{(n)}\(w\)$ given in \rf{fndef}. 

\vspace{1cm}

\noindent {\bf Acknowledgements:} ACRB is supported by a CNPq
scholarship, and LAF is partially supported by a CNPq grant.

\end{document}